\documentclass[12pt]{article}
\pdfoutput=1
\usepackage{amsmath}
\usepackage{amssymb}
\usepackage{graphicx}
\usepackage{subcaption}
\usepackage{url}

\usepackage[sort&compress, numbers, merge]{natbib}

\newcommand{\Slash}[1]{{\ooalign{\hfil/\hfil\crcr$#1$}}}

\usepackage[colorlinks=true, linkcolor=blue, citecolor=blue,
urlcolor=black]{hyperref}

\begin{document}
\baselineskip 0.6cm

\def\simgt{\mathrel{\lower2.5pt\vbox{\lineskip=0pt\baselineskip=0pt
           \hbox{$>$}\hbox{$\sim$}}}}
\def\simlt{\mathrel{\lower2.5pt\vbox{\lineskip=0pt\baselineskip=0pt
           \hbox{$<$}\hbox{$\sim$}}}}
\def\simprop{\mathrel{\lower3.0pt\vbox{\lineskip=1.0pt\baselineskip=0pt
             \hbox{$\propto$}\hbox{$\sim$}}}}
\def\tr{\mathop{\rm tr}}

\begin{titlepage}

\begin{flushright}
DESY 15-103 \\
FTPI-MINN-15/33 \\
IPMU15-0098
\end{flushright}

\vskip 1.1cm

\begin{center}

{\Large \bf 
Probing Bino-Wino Coannihilation at the LHC
}

\vskip 1.2cm

Natsumi Nagata${}^{1}$,
Hidetoshi Otono${}^{2}$, 
and
Satoshi Shirai${}^{3}$
\vskip 0.5cm

{\it
$^1$ William I. Fine Theoretical Physics Institute, School of
Physics and Astronomy, \\ 
University of Minnesota, Minneapolis, MN 55455, USA,\\ 
and 
Kavli Institute for the Physics and Mathematics of the Universe (WPI),
 \\The University of Tokyo Institutes for Advanced Study, The University
 of Tokyo, Kashiwa 277-8583, Japan\\ [5pt] 
$^2$ Research Center for Advanced Particle Physics, Kyushu University,
 Fukuoka 812-8581, Japan\\ 
$^3$ {Deutsches Elektronen-Synchrotron (DESY), 22607 Hamburg, Germany}
}

\vskip 1.0cm

\abstract{
 We study bino-wino coannihilation scenario in the so-called spread or
 mini-split supersymmetry. We show that, in this model, a neutral wino
 has a macroscopic decay length in a wide range of parameter
 space. This characteristic feature could be observed as a displaced
 vertex plus missing transverse energy event at the LHC. In this paper, we
 study the current constraints and future prospects on the scenario
 based on the displaced vertex search performed by the ATLAS
 collaboration. It is found that a sizable parameter region can be
 probed at the 8~TeV LHC run. This search strategy will
 considerably extend its reach at the next stage of the LHC running, and
 thus play a crucial role to examine a possibility of bino dark matter
 in the mini-split type supersymmetric models.  
}

\end{center}
\end{titlepage}

\section{Introduction}
\label{sec:intro}

Split supersymmetry (SUSY)~\cite{Wells:2003tf,*Wells:2004di,
ArkaniHamed:2004fb,*Giudice:2004tc,*ArkaniHamed:2004yi,*ArkaniHamed:2005yv}
is an interesting candidate for physics beyond the Standard Model
(SM). This class of SUSY models have a mass spectrum in which scalar
particles except the SM-like Higgs boson have masses much larger than
the electroweak (EW) scale, while gauginos, probably Higgsino as
well, lie not far from the EW scale. With such heavy scalars, the
split SUSY can overcome disadvantages of the weak-scale SUSY, such as
SUSY flavor/CP problems~\cite{Gabbiani:1996hi} and too rapid proton
decay in the minimal SUSY grand unified theory (GUT) \cite{Goto:1998qg,
Murayama:2001ur}. The light fermionic SUSY partners enable the model to
take over advantages of the weak-scale SUSY: the lightest SUSY particle
(LSP) as a dark matter (DM) candidate and gauge coupling unification. 
Thus, although the origin of the EW scale is not easily elucidated in this
model, its phenomenological aspects are quite appealing.

After the discovery of the SM-like Higgs boson
\cite{Aad:2012tfa,*Chatrchyan:2012ufa} with a mass of around 125 GeV
\cite{Aad:2015zhl}, this framework becomes increasingly convincing. 
Especially, a spectrum with mild splitting ~\cite{Hall:2011jd,
Hall:2012zp, Ibe:2011aa, *Ibe:2012hu, Arvanitaki:2012ps,
ArkaniHamed:2012gw, Evans:2013lpa}, where the scalar mass scale falls
into from several tens of TeV to a PeV and the gaugino mass scale is
${\cal O}(1)$~TeV, is getting more popular---this
mass spectrum is often called spread or mini-split SUSY, as the 
125~GeV Higgs mass is a sweet spot of this spectrum
\cite{Okada:1990vk,*Okada:1990gg,Ellis:1990nz,
Haber:1990aw,Ellis:1991zd}. Such models can also improve gauge
coupling unification \cite{Hisano:2013cqa}, and accommodate a simple GUT
without suffering from the rapid proton decay problem
\cite{Hisano:2013exa, Nagata:2013sba, Evans:2015bxa}. Moreover,
reasonable assumptions on the multiverse lead to this
spectrum~\cite{Nomura:2014asa}.

The realization of such a mass spectrum is rather easy.
We may simply assume some charge on the SUSY breaking field $X$.
Then, the symmetry associated with the charge forbids Lagrangian terms
$[XW^{\alpha} W_{\alpha}/M_*]_{\theta^2}$, where $W^\alpha$ is a gauge
field strength superfield and $M_*$ is a cut-off scale,
\textit{e.g.}, the Planck scale. Since these terms reduce to the gaugino
mass terms after the SUSY breaking, absence of these terms implies
suppression of gaugino masses.
On the other hand, terms like $[XX^{\dagger} \Phi \Phi^{\dagger} /
M_*^2]_{\theta^4}$, where $\Phi$ is a matter chiral superfield, are
generally allowed and lead to soft scalar masses of $\widetilde{m} = {\cal
O}(F_X/M_*)$, with $F_X$ the $F$-term of the
SUSY breaking field $X$. The Higgsino mass $\mu$ is usually expected to
be of the same order of sfermion masses: $|\mu| \sim \widetilde{m}$,
though there are several models that predict a smaller value for
$|\mu|$. Gaugino masses may come from the anomaly mediation
mechanism~\cite{Randall:1998uk, Giudice:1998xp} as well as threshold effects of
the Higgs fields~\cite{Pierce:1996zz} or extra matter
fields~\cite{Pomarol:1999ie,*Nelson:2002sa, *Hsieh:2006ig,*Gupta:2012gu,
*Nakayama:2013uta,*Harigaya:2013asa, *Evans:2014xpa}. 
In these cases, the gaugino masses are suppressed by a loop factor
compared to the scalar mass scale $\widetilde{m}$, and thus the
spread/mini-split spectrum can be realized. In addition, this
hierarchical mass spectrum makes the mixing among gauginos and Higgsino
negligible, and thus they can be regarded as almost pure states.

If the contribution from the anomaly mediation is dominant, the LSP is a
pure wino.
This wino LSP has various interesting features as a DM candidate.
Since winos are charged under the EW
interactions, their self-annihilation cross sections are rather large,
which allows a wino with a mass up to 3~TeV to be consistent with the
current observed DM density~\cite{Hisano:2006nn}. Such a large
annihilation cross section also makes the indirect detection of wino DM
through cosmic ray signals quite promising
\cite{Hisano:2003ec,*Hisano:2004ds, Cohen:2013ama, Fan:2013faa, Ibe:2015tma, Hamaguchi:2015wga}. Direct
detection of wino DM has also been intensively studied
\cite{Hisano:2010fy, *Hisano:2010ct, *Hisano:2011cs, *Hisano:2012wm,
*Hisano:2015rsa}.  
In addition, a pure wino offers a unique collider signal.
In the
spread/mini-split SUSY scenario, the mass splitting between the neutral
and charged winos is predicted to be fairly small. This renders the
charged wino live long, with a decay length of $c\tau_{\widetilde{W}^0}={\cal O}(1)$~cm.
At collider, a produced charged wino leaves a charged track of this
length, which is very useful for the discovery and measurement of the wino
LSP~\cite{Ibe:2006de,*Asai:2008sk,*Asai:2008im}.

On the other hand, when other contributions like threshold effects are
comparable to that of the anomaly mediation, it is questionable
whether the wino LSP is the case or not. Actually, it turns out that an
$M_*$ which is slightly smaller than the reduced Planck scale easily
leads to the bino LSP~\cite{Hall:2012zp,Nomura:2014asa}. The presence of
extra matters may also favor the bino LSP, since a wino tends to receive
larger quantum corrections from the extra matters due to its larger
gauge coupling compared to a bino. The bino LSP is, however, often
disfavored on the basis of cosmology; on the assumption of the $R$-parity
conservation and conventional cosmological history, the bino LSP case
usually suffers from the overproduction of DM because of its small
self-annihilation cross section. For the bino abundance to be consistent
with the current observation, we need to rely on some exceptional
situation \cite{Griest:1990kh}: gaugino coannihilation or Higgs funnel.
If the Higgsino mass is larger than ${\cal O}(10)$ TeV, the Higgs funnel
cannot work effectively since the bino-Higgs coupling is highly
suppressed. In this case, gaugino coannihilation is the only
possibility. We previously studied the coannihilation of the bino LSP
with a gluino~\cite{Nagata:2015hha}. In this work, we focus on the
bino-wino coannihilation~\cite{Baer:2005jq, ArkaniHamed:2006mb,
Ibe:2013pua, Harigaya:2014dwa}.

The bino DM is quite sterile compared to the wino DM, since both the
self-annihilation cross section and the direct detection rate are
suppressed by heavy masses of sfermions and Higgsino. Thus, probing this
spectrum with the DM experiments is extremely challenging. Instead, the bino-wino
coannihilation scenario has a specific mass spectrum; wino should have a
mass fairly close to the bino mass in order to make coannihilation
effective and to assure that the relic abundance of the bino DM is less
than or equal to the observed DM density. Previous works have revealed
that the mass difference $\Delta M$ should be $\lesssim {\cal
O}(10)$~GeV \cite{Baer:2005jq, ArkaniHamed:2006mb, Ibe:2013pua,
Harigaya:2014dwa} to satisfy the condition. Such a small mass difference
makes it possible to probe the scenario in collider experiments, since a
degenerate mass spectrum often gives rise to a long-lived particle,
which offers a distinct signature. Indeed, we find that a neutral
wino can actually be long-lived in our setup and thus be a nice target
to probe the scenario.

In this work, we study in detail the decay of a neutral wino in the
bino-wino coannihilation scenario and show that it typically has
a decay length larger than ${\cal O}(1)$~mm. Then, we consider the
detectability of the neutral wino decay at the LHC. It is found that
searches for a displaced vertex (DV) can actually be a powerful probe
for a neutral wino with such a long decay length, and thus provide a
promising way of testing the bino-wino coannihilation scenario.

This paper is organized as follows. In the next section, we formulate an
effective theory for bino and wino to study the wino decay in the
spread/mini-split spectrum. Then, in  Sec.~\ref{sec:lhcsearch}, we
discuss the current constraints and future prospects on the searches for
the decay signature of a long-lived neutral wino. Finally,
Sec.~\ref{sec:conclusion} is devoted to conclusion and discussion.

\section{Wino decay}

\begin{figure}[t!]
\centering
\includegraphics[clip, width = 0.6 \textwidth]{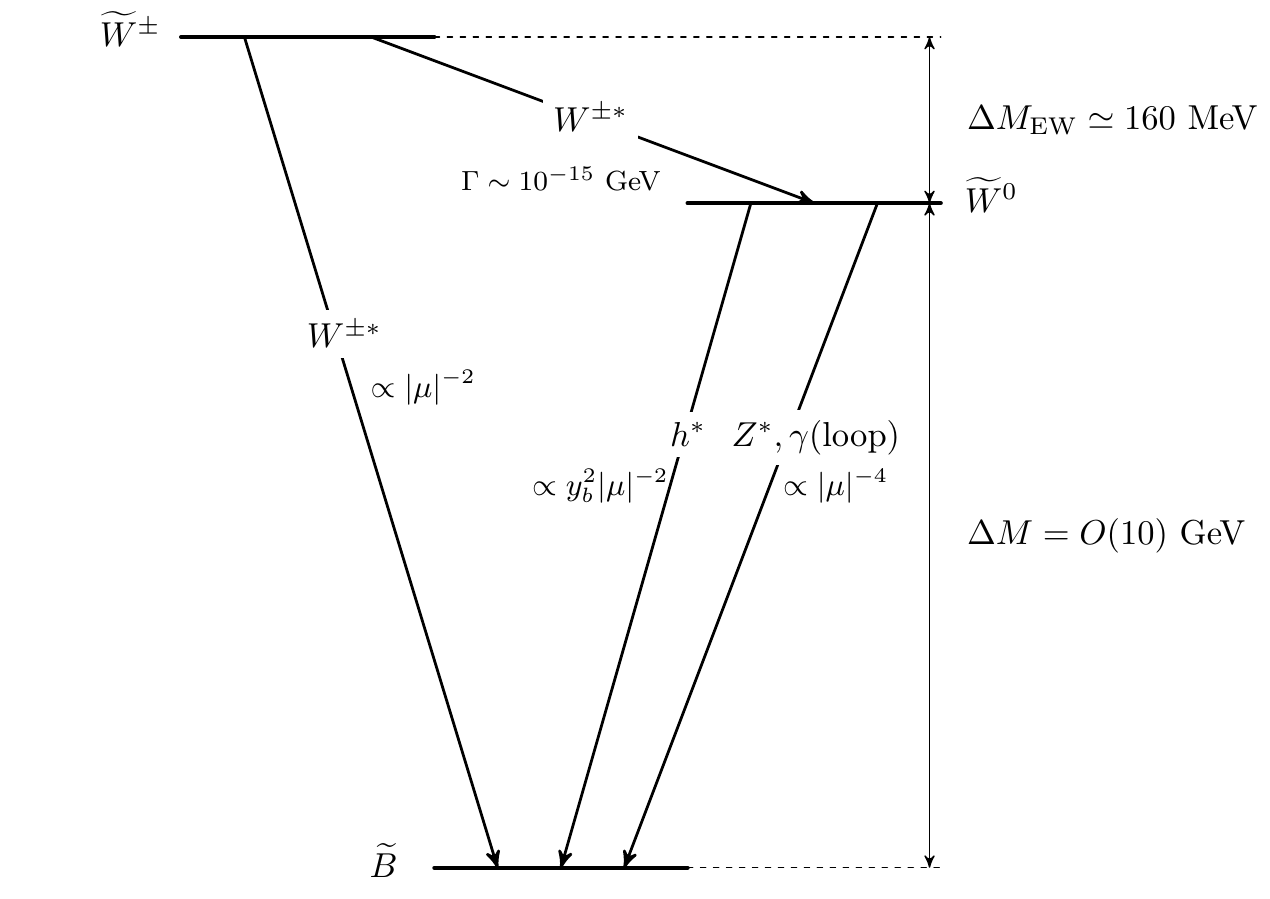}
\caption{Mass spectrum and decay chains of the present model.}
\label{fig:spectrum}
\end{figure}

In this section, we discuss the decay properties of a neutral wino in
the case where the neutral wino is highly degenerate with the bino LSP
in mass and Higgsino and sfermions are much heavier than these EW
gauginos. To adequately deal with this hierarchical setup, in
Sec.~\ref{sec:effectivethe}, we first construct a low-energy effective
theory for the EW gauginos by integrating out these heavy particles. Then, in
Sec.~\ref{sec:decayl}, we discuss the decay properties of the neutral
wino. We also show expected values for its decay length taking into
account the thermal relic abundance of the bino LSP.

Before going into the detailed discussion, let us first summarize the
results of this section. In Fig.~\ref{fig:spectrum}, we show
the mass spectrum for the EW gauginos and the suppression factors in
their decay rates. We assume the bino-wino mass difference, $\Delta M$,
to be ${\cal O}(10)$~GeV, which leads to successful DM
coannihilation~\cite{Baer:2005jq, ArkaniHamed:2006mb, Ibe:2013pua,
Harigaya:2014dwa}. The decay rate of wino into the bino LSP is
suppressed by the heavy Higgsino mass. Charged wino can decay into bino
promptly for $|\mu|< {\cal O}(10)$~PeV, since the
decay occurs via a dimension-five operator. On the other hand, the
neutral wino decay is not so rapid. If $\Delta M$ is less than the $Z$
boson mass, neutral wino decays into bino only through the virtual
$Z$ boson or Higgs boson $h$ exchange, or via the two-body decay process
with emitting a photon at loop level. As we see below, the $Z$ boson
mediated decay and the two-body photon-emitting processes are suppressed
by a factor of $|\mu|^{-4}$. Regarding the Higgs boson mediated decay,
on the other hand, its decay rate is only suppressed by a factor of
$|\mu|^{-2}$, though the small couplings
between the Higgs boson and the SM fermions prevent neutral wino from
decaying rapidly. As a result, for $|\mu|\gtrsim 10$ TeV, the decay
length $c\tau_{\widetilde{W}^0}$ of neutral wino gets macroscopic: $c\tau_{\widetilde{W}^0} \gtrsim {\cal
O}(1)$~mm. 

\subsection{Effective theory for wino decay}
\label{sec:effectivethe}

Here, we discuss the wino decay based on the effective field theoretical
approach. To begin with, we introduce the full theory containing
Higgsino with renormalizable interactions. Then, we obtain a relevant
effective field theory by integrating out the heavy Higgsino. The
Higgsino contributions are described by higher-dimensional operators in
the effective theory, which causes the wino decay into the bino LSP. 

\subsubsection*{Full theory above Higgsino scale}

First, let us consider the full theory. In the mini-spit/spread SUSY, it
is reasonable to assume that the sfermion mass scale $\widetilde{m}$ is
similar to or greater than the Higgsino mass $\mu$:  $\widetilde{m}
\gtrsim |\mu|$. In this case, the decay of wino is dominantly controlled
by the gaugino-Higgsino-Higgs couplings, rather than the interactions
with sfermions and heavy Higgs bosons,
and thus we can safely neglect their contributions in the following
discussion.

The gaugino-Higgsino-Higgs interactions are given by
\begin{align}
{\cal L}_{\text{int}} =&  -\frac{1}{\sqrt{2}}
\{g_{1u}^{} H^{\dagger} \widetilde{H}_u+
 g_{1d}^{} \epsilon^{\alpha\beta} (H)_\alpha(\widetilde {H}_d)_\beta
 \} \widetilde{B} \nonumber \\
  &-
  {\sqrt{2}}\{
g_{2u}^{} H^{\dagger}T^A \tilde H_u 
 -g_{2d}^{} \epsilon^{\alpha\beta}(H)_\alpha (T^A \widetilde{H}_d)_\beta
  \}\widetilde{W}^A +\text{h.c.}~,
\label{eq:gauginohiggsinocoup}
\end{align}
where $\widetilde{H}_{u,d}$, $\widetilde{B}$, and $\widetilde{W}^A$
($A=1,2,3$) denote the Higgsino, bino, and wino fields, respectively;
$H$ is the SM Higgs field; $T^A$ are the SU(2)$_L$ generators.
In this paper, we mainly use the two-component notation for fermion
fields unless otherwise noted.
At the leading order, the above coupling constants are given by
\begin{align}
g_{1u}^{} &= g^\prime \sin\beta, ~~~~~~g_{1d}^{} = g^\prime \cos\beta~, \nonumber\\
g_{2u}^{} &= g \sin\beta, ~~~~~~~g_{2d}^{} = g \cos\beta~,
\label{eq:gauginocoupmatch}
\end{align}
at the SUSY breaking scale $\widetilde{m}$. Here, $g^\prime$ and $g$ are
the U(1)$_Y$ and SU(2)$_L$ gauge coupling constants, respectively, and
$\tan\beta \equiv \langle H_u^0\rangle /\langle H_d^0\rangle$. 

The gaugino and Higgsino mass terms are defined by
\begin{align}
{\cal L}_{\text{mass}}=
-\frac{M_1}{2}\widetilde{B}\widetilde{B}
-\frac{M_2}{2}\widetilde{W}^A \widetilde{W}^A
-\mu ~\epsilon^{\alpha\beta}(\widetilde{H}_u)_\alpha
 (\widetilde{H}_d)_\beta 
+\text{h.c.}~,
\end{align}
with $\epsilon^{\alpha\beta}$ the antisymmetric tensor. 
In the following subsection, we construct an effective filed theory for
the gauginos by integrating out the Higgsinos $\widetilde{H}_u$ and
$\widetilde{H}_d$, which are supposed to be much heavier than the gauginos.

\subsubsection*{Effective theory below Higgsino scale}

Next, we formulate an effective theory which describes the
wino decay into the bino LSP. The decay is caused by effective
interactions expressed by higher-dimensional operators, which are
induced when we integrate out heavier particles than
gauginos---Higgsinos and scalar particles whose masses are ${\cal
O}(10$--$10^3)$~TeV. Let us write down relevant operators up to
dimension six. For the dimension-five operators, we have 
\begin{align}
 {\cal O}_1^{(5)} &= \widetilde{B}\widetilde{W}^A
H^\dagger T^A H ~, \\
 {\cal O}_2^{(5)} &= \widetilde{B}\sigma^{\mu\nu} \widetilde{W}^A
W^A_{\mu\nu} ~, \\
 {\cal Q}_1 &= \frac{1}{2}{\widetilde{B}}\widetilde{B} |H|^2 ~,
  \\
 {\cal Q}_2 &= \frac{1}{2}{\widetilde{W}^A}\widetilde{W}^A |H|^2 ~,
\end{align}
where  $W^A_{\mu\nu}$ is the SU(2)$_L$ gauge field
strength tensor; $\sigma_{\mu\nu}\equiv \frac{i}{2}(\sigma_\mu
\overline{\sigma}_\nu -\sigma_\nu \overline{\sigma}_\mu)$, where
$\sigma^\mu =(\sigma^0, \sigma^i)$ and $\overline{\sigma}^\mu
=(\sigma^0, -\sigma^i)$ with $\sigma^i$ ($i=1,2,3$) the Pauli matrices.  
 The first two operators contain a bino and a wino, and thus directly contribute to the wino decay into the bino
LSP. The latter two are, on the other hand, only relevant to the mass
matrix for the neutral bino and wino; these operators reduce to the mass
terms for them after the EW symmetry breaking. As for
dimension-six, we have 
\begin{align}
{\cal O}^{(6)} &= {\widetilde{B}}^\dagger \overline{\sigma}^\mu
 \widetilde{W}^A H^\dagger T^A i\overleftrightarrow{D}_\mu H ~, 
\end{align}
where $D_\mu$ is the covariant derivative and $A\overleftrightarrow{D}_\mu B
\equiv AD_\mu B - (D_\mu B) A$ with $A$ and $B$ arbitrary fields. This
operator also contributes to 
the wino decay, though its effect is further suppressed by a heavy
mass scale. Then, the effective interactions are given as follows:
\begin{equation}
 \Delta {\cal L}_{\text{int}} = \sum_{i=1,2} C_i^{(5)} {\cal O}^{(5)}_i +
\sum_{i=1,2}\widetilde{C}_i {\cal Q}_i + 
C^{(6)} {\cal O}^{(6)} +
\text{h.c.}
\end{equation}
The Wilson coefficients of these operators are determined below.

\begin{figure}[t]
\centering
\subcaptionbox{\label{fig:higgsino_tree} Tree level}{\includegraphics[width=0.39\textwidth]{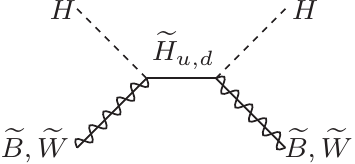}}
\hspace{1cm}
\subcaptionbox{\label{fig:higgsino_loop} One loop level}{\includegraphics[width=0.35\textwidth]{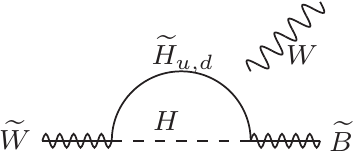}} 
\caption{
Examples of the diagrams, which generate the effective operators.
}
\label{fig:higgsino}
\end{figure}

By evaluating the tree-level Higgsino exchange diagrams (Fig.~\ref{fig:higgsino_tree}), we readily obtain
\begin{align}
 C^{(5)}_1 &= \frac{1}{\mu}(g_{1u}^{}g_{2d}^{}+g_{1d}^{} g_{2u}^{})
+\frac{1}{2|\mu|^2}[(g_{1u}^* g_{2u}^{}+g_{1d}^*g_{2d}^{})M_1+
(g_{1u}^{} g_{2u}^*+g_{1d}^{}g_{2d}^*)M_2]
~, \label{eq:c51}\\
C^{(6)} &= -\frac{1}{2|\mu|^2}(g_{1u}^*g_{2u}^{}-g_{1d}^* g_{2d}^{}) ~,\\
 \widetilde{C}_1 &= \frac{g_{1u}^{}g_{1d}^{}}{\mu} +\frac{M_1}{2|\mu|^2}
 (|g_{1u}^{}|^2 + |g_{1d}^{}|^2) ~, \\
 \widetilde{C}_2 &= \frac{g_{2u}^{}g_{2d}^{}}{\mu} +\frac{M_2}{2|\mu|^2}
 (|g_{2u}^{}|^2 + |g_{2d}^{}|^2) ~. 
\end{align}
Here we have kept effective operators up to dimension six, and used
equations of motions for external gaugino fields to eliminate redundant
operators. The operator ${\cal O}^{(5)}_2$ is not induced at tree
level. However, since this operator gives rise to the two-body
decay process $\widetilde{W}^0 \to \widetilde{B} + \gamma$, it
could be important even though it is induced at loop level
\cite{Haber:1988px, Ambrosanio:1996gz, Diaz:2009zh, Han:2014xoa},
especially when the wino mass is close to the bino mass. Thus, only
for this operator, we also consider the one-loop contribution. The
one-loop Higgsino-Higgs loop diagram (Fig.~\ref{fig:higgsino_loop}) yields 
\begin{align}
 C_2^{(5)} =& +\frac{g}{2(4\pi)^2 \mu} (g_{1u}^{}g_{2d}^{}-g_{1d}^{}g_{2u}^{}) 
\nonumber \\
&-\frac{g}{8(4\pi)^2}\left[
(g_{1u}^*g_{2u}^{}-g_{1d}^*g_{2d}^{})\frac{M_1}{|\mu|^2}
-(g_{1u}^{}g_{2u}^*-g_{1d}^{}g_{2d}^*)\frac{M_2}{|\mu|^2}
\right] ~,
\end{align}
where again we have kept terms up to ${\cal O}(|\mu|^{-2})$. 
Note that the first term vanishes if we use the tree-level relation
Eq.~\eqref{eq:gauginocoupmatch}.  We also find that the heavy Higgs
contribution of ${\cal O}(\mu^{-1})$ vanishes in a similar manner. Thus, although the 
operator ${\cal O}^{(5)}_2$ is dimension five, its Wilson coefficient is
suppressed by $|\mu|^{-2}$ and thus subdominant compared to the
contribution of ${\cal O}^{(5)}_1$. Moreover, the terms in the second
line could also cancel with each other to great extent if $M_1\simeq
M_2$. This results in a further suppression of this contribution. 
Besides, quark-squark loop processes can generate the operator ${\cal
O}^{(5)}_2$ at one-loop level. Their contribution is
suppressed by a factor of $m_{\tilde{q}}^{-2}$ on top
of a loop factor, with $m_{\widetilde{q}}$ the mass of the
squark running in the loop, and thus again subdominant.

\subsubsection*{EW broken phase}

After the Higgs field acquires a vacuum expectation value (VEV), the
operators ${\cal O}^{(5)}_1$, ${\cal Q}_1$, and ${\cal Q}_2$ reduce to
the mass terms for bino and wino. The mass matrix for
the neutral sector is given by
\begin{equation}
 {\cal L}_{\text{mass}} =-\frac{1}{2}
(\widetilde{B}~\widetilde{W}^0) {\cal M}
\begin{pmatrix}
 \widetilde{B} \\ \widetilde{W}^0
\end{pmatrix}
~,
\end{equation}
with 
\begin{equation}
 {\cal M}=
\begin{pmatrix}
 M_1-\frac{v^2}{2}\widetilde{C}_1 & \frac{v^2}{4}C_1^{(5)} \\
 \frac{v^2}{4}C_1^{(5)} & M_2 -\frac{v^2}{2}\widetilde{C}_2
\end{pmatrix}
\equiv 
\begin{pmatrix}
 {\cal M}_{11} & {\cal M}_{12} \\ {\cal M}_{12} & {\cal M}_{22}
\end{pmatrix}
~,
\end{equation}
where $v\simeq 246$~GeV is the Higgs VEV. 
This mass matrix can be diagonalized with a $2\times 2$ unitary matrix
$U$, which we parametrize by 
\begin{align}
 U=
\begin{pmatrix}
 e^{i\alpha}& 0 \\ 0& e^{i\beta}
\end{pmatrix}
\begin{pmatrix}
 \cos\theta & e^{-i\phi}\sin\theta \\
 -e^{i\phi} \sin\theta & \cos\theta
\end{pmatrix}
~.
\end{align}
Then, we find that the matrix $M$ is diagonalized as \cite{Takagi,Choi:2006fz}
\begin{equation}
 U^*{\cal M}U^\dagger =
\begin{pmatrix}
 m_1&0\\ 0 & m_2
\end{pmatrix}
~,
\end{equation}
where $m_1$ and $m_2$ are real and non-negative, whose values are given
by
\begin{equation}
   m_{1,2}^2 =
\frac{1}{2}[|{\cal M}_{11}|^2+|{\cal M}_{22}|^2+2|{\cal M}_{12}|^2\mp
\sqrt{(|{\cal M}_{11}|^2-|{\cal M}_{22}|^2)^2
+4|{\cal M}_{11}^*{\cal M}_{12}+{\cal M}_{22}{\cal M}_{12}^*|^2}]~.
\end{equation}
The mixing angle $\theta$ in the unitary matrix $U$ is given by
\begin{equation}
 \tan\theta =\frac{|{\cal M}_{11}|^2-|{\cal M}_{22}|^2+\sqrt{(|{\cal
 M}_{11}|^2-|{\cal M}_{22}|^2)^2 
+4|{\cal M}_{11}^*{\cal M}_{12}+{\cal M}_{22}{\cal M}_{12}^*|^2}}
{2|{\cal M}_{11}^*{\cal M}_{12}+{\cal M}_{22}{\cal M}_{12}^*|}~, 
\end{equation}
while its phase factors are 
\begin{align}
  e^{i\phi}&=\frac{{\cal M}_{11}^*{\cal M}_{12}+{\cal M}_{22}{\cal
 M}_{12}^*} {|{\cal M}_{11}^*{\cal M}_{12}+{\cal M}_{22}{\cal M}_{12}^*|}~, \\
 \alpha &= \frac{1}{2}\arg\bigl({\cal M}_{11}-{\cal M}_{12}e^{-i\phi} 
\tan\theta\bigr)~,\\
 \beta &= \frac{1}{2}\arg\bigl({\cal M}_{22}+{\cal
 M}_{12}e^{i\phi}\tan\theta\bigr)~. 
\end{align}
In this calculation, we have implicitly assumed that $|{\cal
M}_{11}|\leq |{\cal M}_{22}|$. 
 In terms of the gaugino masses and the  
Wilson coefficients, these parameters are approximately given as
\begin{align}
 m_1^2 &\simeq |M_1|^2 -v^2 \text{Re}(M_1\widetilde{C}^*_1) ~,
\nonumber \\
 m_2^2 &\simeq |M_2|^2 -v^2 \text{Re}(M_2\widetilde{C}^*_2) ~,
\nonumber \\
 \tan\theta &\simeq \frac{v^2}{4}\frac{|M_1^* C_1^{(5)}+M_2C_1^{(5)*}|}
{|M_2|^2-|M_1|^2}  ~, \nonumber \\
 \phi &\simeq \text{arg}(M_1^*C_1^{(5)}+M_2 C_1^{(5)*}) ~, \nonumber \\
 \alpha &\simeq
\frac{1}{2}\text{arg}(M_1)
 -\frac{v^2}{4}\text{Im}\left(\frac{\widetilde{C}_1}{M_1}\right) ~, 
\nonumber \\
 \beta &\simeq \frac{1}{2}\text{arg}(M_2)
 -\frac{v^2}{4}\text{Im}\left(\frac{\widetilde{C}_2}{M_2}\right) ~.
\label{eq:paramapp}
\end{align}
If $|M_2|-|M_1|$ is ${\cal O}(10)$~GeV, which is motivated by the
bino-wino coannihilation scenario as shown in Refs.~\cite{Baer:2005jq,
ArkaniHamed:2006mb, Ibe:2013pua, Harigaya:2014dwa}, then the above
approximations are valid when $|\mu|\gtrsim {\cal O}(10)$~TeV, with which the
bino-wino mixing angle is sufficiently small: $\tan\theta \ll 1$. The
mass eigenstates are related to the weak eigenstates through the unitary
matrix $U$ by
\begin{equation}
 \begin{pmatrix}
  \widetilde{\chi}^0_1 \\  \widetilde{\chi}^0_2
 \end{pmatrix}
=
U
\begin{pmatrix}
 \widetilde{B} \\ \widetilde{W}^0
\end{pmatrix}
~.
\end{equation}
In this paper, we assume the bino-like state $\widetilde{\chi}^0_1$ is slightly
lighter than the neutral wino-like state $\widetilde{\chi}^0_2$: $m_1
\lesssim m_2$ with the mass difference $\Delta M \equiv m_2-m_1$ being
${\cal O}(10)$~GeV. 

Next, we consider chargino $\widetilde{\chi}^+$, which is related to the
weak eigenstate by
\begin{equation}
 \widetilde{\chi}^+ =  \frac{e^{i\gamma}}{\sqrt{2}} 
(\widetilde{W}^1-i\widetilde{W}^2)~.
\end{equation}
Its mass eigenvalue and phase factor $\gamma$ are given by
\begin{align}
 m_{\widetilde{\chi}^+} =\left\vert 
M_2 -\frac{v^2}{2}\widetilde{C}_2\right\vert  ~,
~~~~~~
\gamma \simeq \beta ~.
\label{eq:chargmass}
\end{align}
On top of that, EW loop corrections make the chargino
heavier than the neutral wino by a small amount; this contribution to
the mass splitting is evaluated as
$\Delta M_{\text{EW}} \simeq 160$~MeV at two-loop level
\cite{Yamada:2009ve, Ibe:2012sx}. Note that up
to dimension six the higher-dimensional operators do not generate the
mass difference between the neutral and charged winos, as we have seen
from Eq.~\eqref{eq:paramapp} and Eq.~\eqref{eq:chargmass}. For this reason,
although the EW correction $\Delta M_{\text{EW}}$ is quite
small, it turns out to be  the dominant contribution to the mass
difference as long as $|\mu| \gtrsim 10$~TeV.

Now we summarize the interactions relevant to the decay of
$\widetilde{\chi}^0_2$ and $\widetilde{\chi}^+$. First, we consider the
chargino decay. 
In this scenario, a chargino $\widetilde{\chi}^+$ mainly decays into
$\widetilde{\chi}^0_1$ because of the degeneracy between
$\widetilde{\chi}^0_2$ and $\widetilde{\chi}^+$. This decay is caused by
the tree-level gauge interactions through the bino-wino mixing. In the
mass eigenbasis, the gauge interactions are written as
\begin{align}
 {\cal L}_{\widetilde{\chi}^0 \widetilde{\chi}^+ W} =
-g\sin\theta \overline{\widetilde{\chi}^0_1}\Slash{W}^-
\left[e^{-i(\phi -\alpha +\beta)}P_L+e^{i(\phi -\alpha +\beta)}P_R \right]
 \widetilde{\chi}^+
-g\cos\theta \overline{\widetilde{\chi}^0_2}\Slash{W}^-
 \widetilde{\chi}^+
+\text{h.c.}~,
\label{eq:gaugeint}
\end{align}
where we have used four-component notation. Notice that these interactions
are invariant under the charge conjugation. We use this property
below. Using the tree-level relation (\ref{eq:gauginocoupmatch}), we
can obtain an approximate expression for the
$\widetilde{\chi}^0_1$-$\widetilde{\chi}^{\pm}$-$W^{\mp}$ coupling as
\begin{equation}
 - g\sin\theta \simeq -g \sin 2\beta \frac{m_W^2\tan\theta_W}{\mu \Delta
  M} ~,
\label{eq:approxchichiw}
\end{equation}
where $\theta_W$ is the weak-mixing angle and $m_W$ is the $W$-boson
mass. We take the gaugino masses to be real in the derivation.

Second, we discuss the decay of heavier neutralino
$\widetilde{\chi}^0_2$ into the lightest neutralino
$\widetilde{\chi}^0_1$. This decay process occurs via the (off-shell)
Higgs emission induced by the dimension-five operator ${\cal
O}^{(5)}_1$. The relevant
interaction is given by 
\begin{equation}
 {\cal L}_{\widetilde{\chi}^0_1\widetilde{\chi}^0_2 h }
=-\frac{v}{2}e^{-i(\alpha + \beta)}
\cos 2\theta C_1^{(5)} h{\widetilde{\chi}^0_1}
\widetilde{\chi}^0_2 +\text{h.c.} 
\label{eq:muhiggschi1chi2}
\end{equation} 
Since ${\cal O}^{(5)}_1$ is of dimension five, this interaction is only
suppressed by $|\mu|^{-1}$. Similarly to Eq.~\eqref{eq:approxchichiw},
we can approximate the coupling by $-g\tan\theta_W \sin(2\beta)
m_W/\mu$. Notice that large $\tan\beta$ suppresses this coupling. In
this case, the contributions suppressed by $|\mu|^{-2}$ originating from
the second term in Eq.~\eqref{eq:c51} may dominate this contribution.

The interactions of $\widetilde{\chi}^0_1$ and $\widetilde{\chi}^0_2$
with a photon and a $Z$ boson are, on the other hand, relatively
small. At renormalizable level, there is no such a interaction since 
$Q=Y=T_3=0$ for both of these particles. As discussed above, the
contribution of the dimension-five operator ${\cal O}^{(5)}_2$ is
suppressed by $|\mu|^{-2}$ besides the one loop factor. The
dimension-six operator ${\cal O}^{(6)}$ also gives rise to such
interactions at tree level,
\begin{equation}
 {\cal L}_{\widetilde{\chi}^0_1\widetilde{\chi}^0_2 Z}
= \frac{g_Zv^2}{4}e^{i(\alpha-\beta)}C^{(6)} 
\widetilde{\chi}^{0\dagger}_1 \overline{\sigma}^\mu 
\widetilde{\chi}^0_2 Z_\mu +\text{h.c.}~,
\label{eq:chichiz}
\end{equation}
where $g_Z \equiv \sqrt{g^{\prime 2}+g^2}$. 
This interaction is again suppressed by
$\cos(2\beta)|\mu|^{-2}$. Consequently, compared to the Higgs interaction in
Eq.~\eqref{eq:muhiggschi1chi2}, the interactions with a photon and a $Z$
boson are fairly small if we take $|\mu|$ to be sufficiently large. 

An important caveat here is that these interactions could be important
if $\tan \beta$ is large, since the Higgs coupling in
Eq.~\eqref{eq:muhiggschi1chi2} is highly suppressed in this case as
mentioned above. Although the mini-split type models
favor small $\tan \beta$ to explain the 125~GeV Higgs mass, a moderate
size of $\tan \beta$ may be allowed if stop masses are rather light. In
such cases, the terms suppressed by $|\mu|^{-2}$ can also be
significant.

\subsection{Decay length in bino-wino coannihilation}
\label{sec:decayl}

\begin{figure}[t]
\centering
\subcaptionbox{\label{fig:h}Higgs
 exchange}{\includegraphics[width=0.39\textwidth]{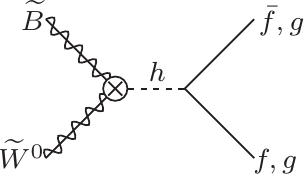}}
\hspace{1cm}
\subcaptionbox{\label{fig:z}$Z$ boson
 exchange}{\includegraphics[width=0.35\textwidth]{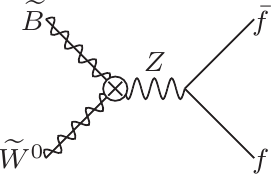}} 
\caption{
Tree-level contributions to the decay of a neutral wino into the bino LSP.
The symbols $\bigotimes$ in the diagrams represent the effective interactions, which are  suppressed by powers of the inverse of the Higgsino mass $\mu$.
}
\label{fig:treewb}
\end{figure}

Taking the above discussion into account, we now estimate the decay
length and branching ratios of the neutral wino-like state
$\widetilde{\chi}^0_2$. As we have seen above, when $|\mu|$ is large
enough, the Higgs interaction \eqref{eq:muhiggschi1chi2} dominates the
others. This interaction gives rise to three-body decay processes
shown in Fig.~\ref{fig:h} when the mass difference between
$\widetilde{\chi}^0_1$ and $\widetilde{\chi}^0_2$, $\Delta M$, is as
small as ${\cal O}(10)$~GeV. In this case, the dominant decay channel is
$\widetilde{\chi}^0_2 \to \widetilde{\chi}^0_1 b \bar{b}$, whose decay
amplitude is suppressed by the bottom-quark Yukawa coupling in addition
to the three-body phase space factor and $|\mu|^{-1}$. Moreover, the
decay rate is kinematically suppressed due to the small mass difference
$\Delta M$. For these reasons, the resultant decay rate is considerably
small and $\widetilde{\chi}^0_2$ has a sizable decay length, as we
see below. 

Notice that, as we mentioned above, tree-level sfermion exchange
contributions to the three-body decay processes are generically
much smaller than the Higgs exchange contribution. The former
contributions can be expressed by dimension-six effective operators whose
coefficients are $\sim gg^\prime /\widetilde{m}^2$. On the other hand,
after integrating out the Higgs boson, the diagram in Fig.~\ref{fig:h}
yields a dimension-six operator with its coefficient being $\sim
gg^\prime (\sin 2\beta) m_b/(\mu m_h^2)$. Therefore, the Higgs exchange
contribution dominates sfermion ones as long as $|\mu| \ll {\cal
O}(10^3)~\text{TeV}\times\sin 2\beta \cdot (\widetilde{m}/10^2~\text{TeV})^2$.

At tree-level, we also have the virtual $Z$ exchange contribution
induced by the interaction \eqref{eq:chichiz}, which is illustrated in
Fig.~\ref{fig:z}. As discussed above, the interaction \eqref{eq:chichiz} is
suppressed by a factor of $|\mu|^{-2}$, and thus their contribution is
subdominant. Notice that ordinary strategies on the searches for charginos
and neutralinos at the LHC rely on the leptonic decay channel
$\widetilde{\chi}^0_2 \to \widetilde{\chi}^0_1 \ell^+ \ell^-$ induced by
this contribution \cite{Aad:2014vma, Khachatryan:2014qwa}. Since the
decay branch into the channel is extremely suppressed in our scenario,
we need an alternative way to probe the bino-wino coannihilation region at
the LHC. This is the subject of the next section.

\begin{figure}[t]
\centering
\subcaptionbox{\label{fig:gamma}Dipole-type contribution
 }{\includegraphics[width=0.30\textwidth]{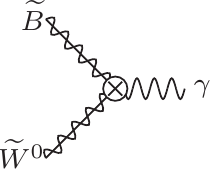}}
\hspace{1cm}
\subcaptionbox{\label{fig:box}Box diagram}{\includegraphics[width=0.43\textwidth]{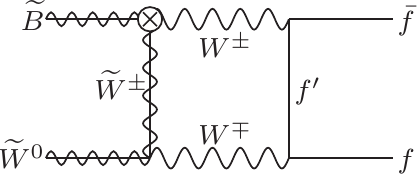}}
\caption{
One-loop contributions to the decay of a neutral wino into the bino LSP.
The symbols $\bigotimes$ in the diagrams represent the effective interactions, which are  suppressed by powers of the inverse of the Higgsino mass $\mu$.
}
\label{fig:loopwb}
\end{figure}

The two-body decay process $\widetilde{\chi}^0_2 \to
\widetilde{\chi}^0_1 \gamma$ shown in Fig.~\ref{fig:gamma} is induced by
the dipole-type operator ${\cal O}^{(5)}_2$, which is generated at
one-loop level. As discussed above, this contribution is also suppressed
by $|\mu|^{-2}$. As a result, it turns out that the three-body decay
processes in Fig.~\ref{fig:h} dominate this two-body process in most of
parameter space we are interested in. One may think that the gauge
interactions \eqref{eq:gaugeint} also induces the process in
Fig.~\ref{fig:gamma} via the chargino-$W$ boson loop diagram. We find, however,
that this contribution vanishes. To see the reason, notice that the
amplitude of the diagram represented in Fig.~\ref{fig:gamma} is odd
under the charge conjugation $C$, as $\widetilde{\chi}^0_1$ and
$\widetilde{\chi}^0_2$ are Majorana fields and a photon is $C$-odd. On
the other hand, as noted above, the gauge interactions in
Eq.~\eqref{eq:gaugeint} preserve the $C$ symmetry. The electromagnetic 
interaction is also invariant under the charge conjugation. Then, it
follows that any amplitude induced by these interactions should be
$C$-even, hence their contribution to the process in
Fig.~\ref{fig:gamma} vanishes. After all, the $\widetilde{\chi}^0_2 \to
\widetilde{\chi}^0_1 \gamma$ decay is sub-dominant as long as $|M_1|, |M_2|,
\ll |\mu|$, and therefore we focus on the tree-level processes in
Fig.~\ref{fig:treewb} in the following analysis.

However, there are several possibilities in which the above conclusion
should be altered. Firstly, if the Higgsino mass is rather small, the
contribution of the dipole operator ${\cal O}^{(5)}_2$ can be
significant. Indeed, such a situation may be realized in the framework
of spread SUSY, as discussed in
Refs.~\cite{Hall:2011jd, Evans:2014pxa, Nagata:2014wma}. If the decay
branch into the $\widetilde{\chi}^0_2 \to \widetilde{\chi}^0_1 \gamma$
channel is sizable, the collider signature discussed in the subsequent section
would be modified. Secondly, as already noted above, the Higgs exchange
contribution decreases if $\tan\beta$ is large. Such a
situation may occur if the SUSY breaking scale is as low as ${\cal
O}(10)$~TeV. Again, in this case, the decay branching ratios may change
considerably. Thirdly, if the mass difference $\Delta M$ is
smaller than ${\cal O}(10)$~GeV, then the $\widetilde{\chi}^0_2 \to
\widetilde{\chi}^0_1 b \bar{b}$ decay mode is highly suppressed, and
other decay modes can be significant, such as one-loop diagrams shown in
Fig.~\ref{fig:box}. In addition, the two-body decay process containing a
bottomonium induced by the virtual Higgs exchange may also be important in
such a situation. Since these possibilities are only relevant to rather
specific parameter region in our setup, we do not consider them in
this paper, and discuss these contributions on another
occasion.

In summary, the mass spectrum for EW gauginos and the
suppression factor in their decay rates are illustrated in
Fig.~\ref{fig:spectrum}. We consider a specific collider signature
for this mass spectrum in the following section.


\begin{figure}[t]
\centering
\subcaptionbox{\label{fig:dl1} $c\tau_{\widetilde{W}^0}$ in $M_{\widetilde{B}}$--$\Delta
 M$ plane.}{\includegraphics[width=0.45\textwidth]{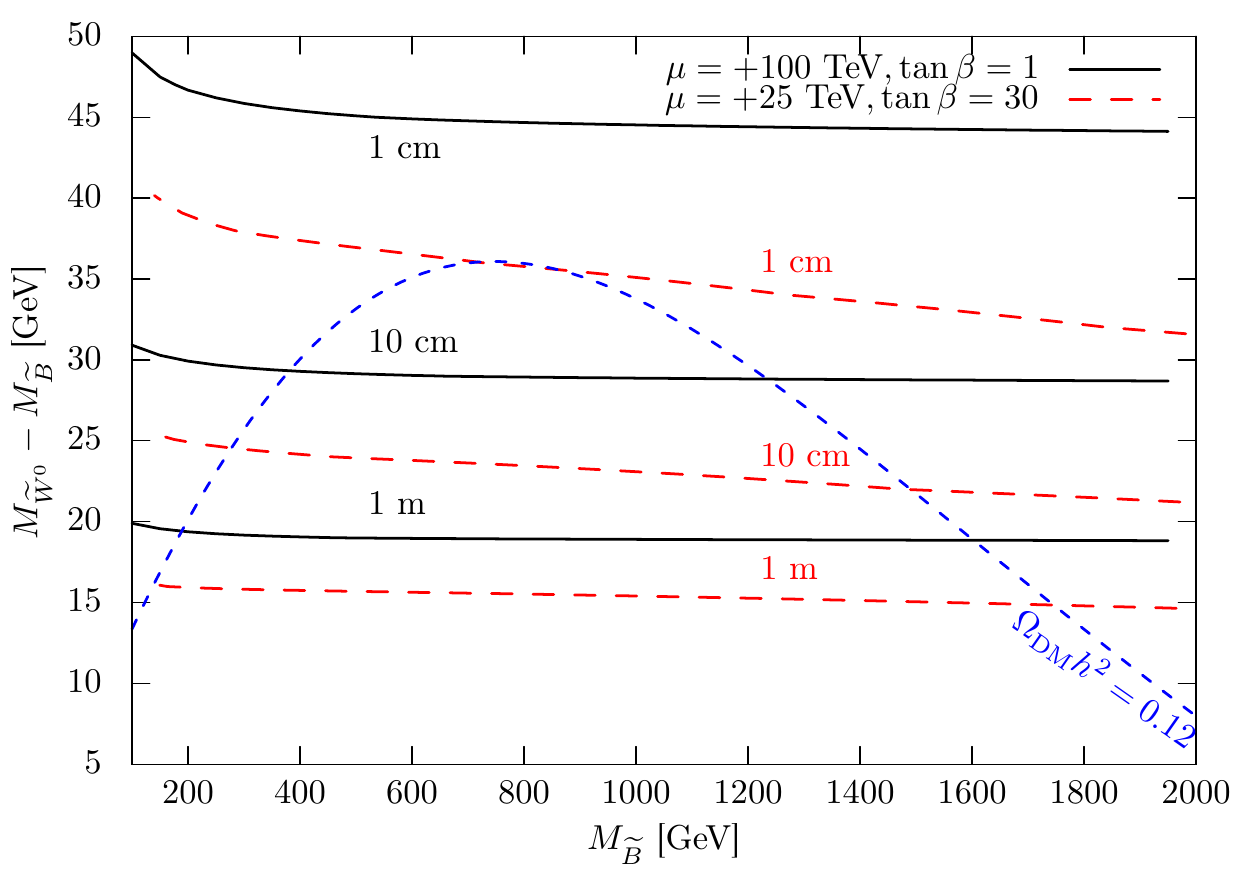}} 
\subcaptionbox{\label{fig:dl2} $c\tau_{\widetilde{W}^0}$ in $\mu$--$\tan\beta$
 plane.}{\includegraphics[width=0.45\textwidth]{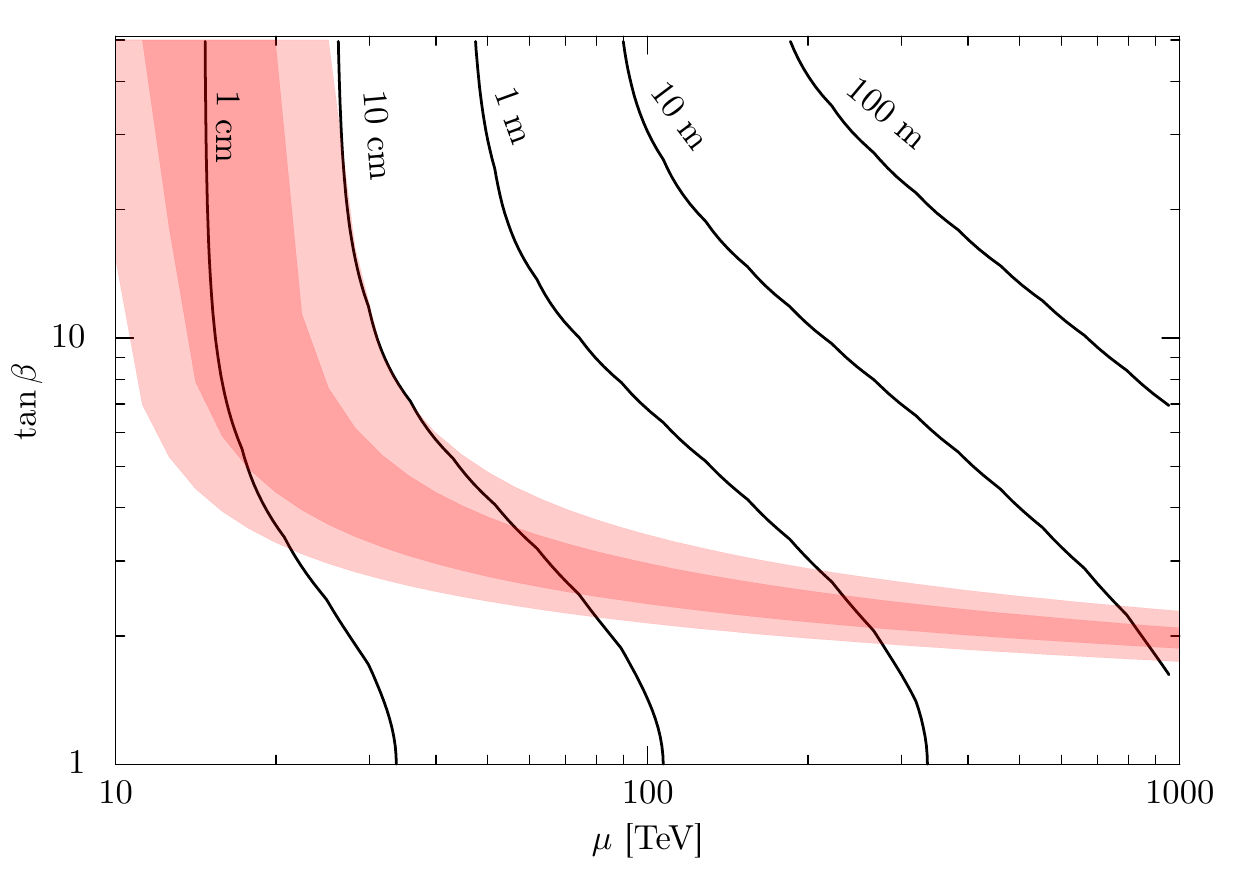}} 
\caption{
Decay length $c \tau_{\widetilde{W}^0}$ of a neutral wino $\widetilde{\chi}^0_2$. 
(a): Black solid (red dashed) lines show contours for the decay length
 of the neutral wino for $\mu=+100$~TeV and $\tan\beta=1$ ($\mu=+25$~TeV and
 $\tan\beta=30$). Blue dashed line corresponds to the parameter region
 which realizes $\Omega_{\text{DM}} h^2 = 0.12$. 
(b): Decay length of the 430 GeV neutral wino into the 400 GeV bino LSP.
The dark (light) red area shows the 1$\sigma$ (2$\sigma$) preferred
 parameter region to explain the 125~GeV Higgs mass. Here we set the
 gluino mass to be 1.5~TeV and the sfermion mass scale to be
 $\widetilde{m} = \mu$.  
}
\label{fig:decay_length}
\end{figure}

Now we evaluate the decay length $c\tau_{\widetilde{W}^0}$ of the neutral wino-like state
$\widetilde{\chi}^0_2$. In Fig.~\ref{fig:dl1}, we plot it in the
$M_{\widetilde{B}}$--$\Delta M$ plane. Here, we set $\mu =+100$~TeV
($+25$~TeV) and $\tan\beta = 1$ (30) in the black solid (red dashed)
lines. The gaugino masses are taken to be real and positive in this
figure. The blue dashed line corresponds to the parameter region where the
thermal relic abundance of the bino LSP agrees to the observed DM
density $\Omega_{\text{DM}} h^2 = 0.12$. As can be seen from this
figure, a neutral wino has a sizable decay length in a wide range of
parameter region motivated by the bino-wino coannihilation scenario.
This observation is a key ingredient for the search strategy discussed
in the next section. In Fig.~\ref{fig:dl2}, on the other hand, the decay
length $c \tau_{\widetilde{W}^0}$ is plotted in the $\mu$--$\tan\beta$ plane. Here, we set
$M_1 = +400$~GeV and $M_2=+430$~GeV. 
For reference, we also show the 1$\sigma$ (2$\sigma$) preferred region
for the 125~GeV Higgs mass in the dark (light) red shaded area, where
the gluino mass is set to be 1.5~TeV and the scalar mass scale is taken
to be equal to the Higgsino mass, $\widetilde{m} = \mu$. 
We find that the decay length grows
as $\tan \beta$ is taken to be large. This is because the Higgs
interaction \eqref{eq:muhiggschi1chi2} is suppressed in this
case. Anyway, a neutral wino has a decay length of $\gtrsim 1$~cm over the
parameter region in this figure, which could be observed at the LHC as
we see in the next section.

Before closing this section, let us comment on the bino-wino coannihilation.
One may wonder the chemical equilibrium between the wino and bino, which
is essential for the successful coannihilation, can be kept even if
$\mu$ is significantly large. Unlike the bino-gluino coannihilation case
\cite{Nagata:2015hha, Ellis:2015vaa}, the bino and wino can interact
with each other through the dimension-five operators. Thus, the interchange
of $\widetilde B \leftrightarrow \widetilde W$ at the freeze-out
temperature is rapid enough for the chemical equilibrium between them to
be maintained, even though $\mu$ is at PeV scale. 

\section{LHC search}
\label{sec:lhcsearch}

We have seen in the previous section that a neutral wino in our scenario
has a sizable decay length. In this section, we discuss the prospects of the
searches for such a long-lived wino at the LHC. A decay process of a
neutral wino with $c \tau_{\widetilde{W}^0} \gtrsim 1$~mm can be observed as
a DV plus missing transverse energy event. The ATLAS collaboration has
reported the result of searches for such events by using the
20.3~fb$^{-1}$ data set collected at the LHC with a center-of-mass
energy of $\sqrt{s}=8$~TeV~\cite{Aad:2015rba}. In
Sec.~\ref{sec:atlasstudy}, we begin with reviewing this ATLAS result for
the DV search. It turns out, however, that the existing result cannot be
directly applied to the bino-wino coannihilation scenario, since the
ATLAS study mainly focus on high-mass DVs. Thus, we need to extend their
analysis so that a smaller mass DV can also be probed. We describe our
setup in Sec.~\ref{sec:setup}. Finally, in Sec.~\ref{sec:prospect},
we show the prospects for the long-lived wino search in the cases of
both direct and gluino mediated productions. 

\subsection{Previous LHC study}
\label{sec:atlasstudy}

The LHC finished the operation at center of mass energy of 8~TeV in 2012.
The ATLAS and the CMS experiments explored many types of neutral
long-lived SUSY particles so far, which are listed as follows
\cite{Aad:2015rba, TheATLAScollaboration:2013yia, Aad:2012zx,
*Aad:2011zb, CMS:2014wda, Aad:2014gfa, CMS:2014hka, Khachatryan:2014mea,
Aad:2015asa, Aad:2015uaa}: 
\begin{itemize}
\item Long-lived gluino in the mini-split SUSY which decays into two
      quarks and a neutralino. 
\item Long-lived neutralino in the $R$-parity violation scenario which
      decays into leptons and quarks. 
\item Long-lived neutralino in the gauge mediation scenario which
      decays into a photon or $Z$ boson, and a gravitino.
\end{itemize}
In these searches, the decay products from the long-lived particles have
large impact parameters, which are distinct from background events that
originate from the primary vertex. Our search strategy proposed in this
paper mainly considers signals including two quarks from a DV, and thus
the long-lived gluino search in the mini-split SUSY performed by the
ATLAS experiment \cite{Aad:2015rba} would be a good reference.   

The ATLAS study \cite{Aad:2015rba} employs several triggers for
long-lived particles. Among them, the trigger utilizing missing
transverse energy ($E^{\rm{miss}}_{\rm{T}}$) has the best sensitivity
for a long-lived gluino; $E^{\rm{miss}}_{\rm{T}} > 80$~GeV is required
for the trigger, $E^{\rm{miss}}_{\rm{T}} > 100$~GeV by offline filters,
and eventually $E^{\rm{miss}}_{\rm{T}} > 180$~GeV at the offline
selection. 

Reconstruction of DVs is CPU-intensive due to many tracks
with high impact parameters. In order to reduce the computation time,
only the events with two ``trackless'' jets of $P_{\rm{T}} >
50$~GeV are processed. The ``trackless'' jets are defined such that the
scalar sum of the transverse momenta of the tracks in the jet should be
less than 5~GeV with the standard track reconstruction. Such jets could
be accompanied by decays of long-lived particles which occur away from
the primary vertex. 

At the end of the selections, a DV with more than four tracks
whose invariant mass is more than 10~GeV is treated as a signal event.
Here, each track is supposed to have the mass of the charged pion
$m_{\pi^{\pm}}$. After the selections, the reconstruction efficiency of
at least a DV from a pair of produced gluinos with a decay length of
$c\tau_{\widetilde{g}}={\cal O}(1)~\rm{mm}$ is almost zero, 
which increases up to more than 50\% when $c\tau_{\widetilde{g}}={\cal O}(10)~\rm{mm}$.
Here, a gluino mass of 1400~GeV is assumed in the estimation. Beyond the
point, the reconstruction efficiency gradually decreases and reaches zero for
$c\tau_{\widetilde{g}} \gtrsim{\cal O}(1)~\rm{m}$. 



In the bino-wino coannihilation scenario, a favored value of $\Delta M$
is $\lesssim 30$~GeV, as can be seen from Fig.~\ref{fig:dl1}. In this
case, jets from a decay of a neutral wino are too soft to satisfy the
$P_\text{T}$ condition  of the trackless jets. 
To probe the scenario with the DV searches, therefore, we need to relax
the above requirements to some extent, which we discuss in what follows.

\subsection{Signal simulation setup}
\label{sec:setup}

Now we adjust the ATLAS DV search method such that it has sensitivity to
long-lived neutral winos, and discuss its prospects at the 14~TeV LHC
running. At the LHC, there are two channels to produce a neutral wino.
One is its direct production and the other is gluino mediated production
in which a produced gluino decays into a neutral wino.
In both cases, the $E^{\rm{miss}}_{\rm{T}}$ trigger is the most
efficient among the DV search triggers, just like the long-lived gluino search. 
In this study, we adopt  $E^{\rm{miss}}_{\rm{T}}>100$ GeV and 200 GeV
for the 8 TeV and 14 TeV LHC cases, respectively.
As for the DV detection criteria, we assume the same setup as the ATLAS study:
\begin{itemize}
\item The number $N_{\rm tr}$ of charged tracks forming a DV should
      be grater than four, where each track should have a transverse
      momentum of $P_{\rm T}>1$~GeV. 
\item The invariant mass of the sum of the momenta of the tracks,
      $m_{\rm DV}$, should be greater than 10~GeV, with the mass of each
      track being assumed to be $m_{\pi^{\pm}}$.
\end{itemize}
Notice that we have dropped the $P_{\text{T}}$ condition for the trackless jets
adopted in the case of the long-lived gluino search.

\begin{figure}[t]
\centering
\subcaptionbox{\label{mdv} Distributions of $m_{\rm
 DV}$}{\includegraphics[width=0.45\textwidth]{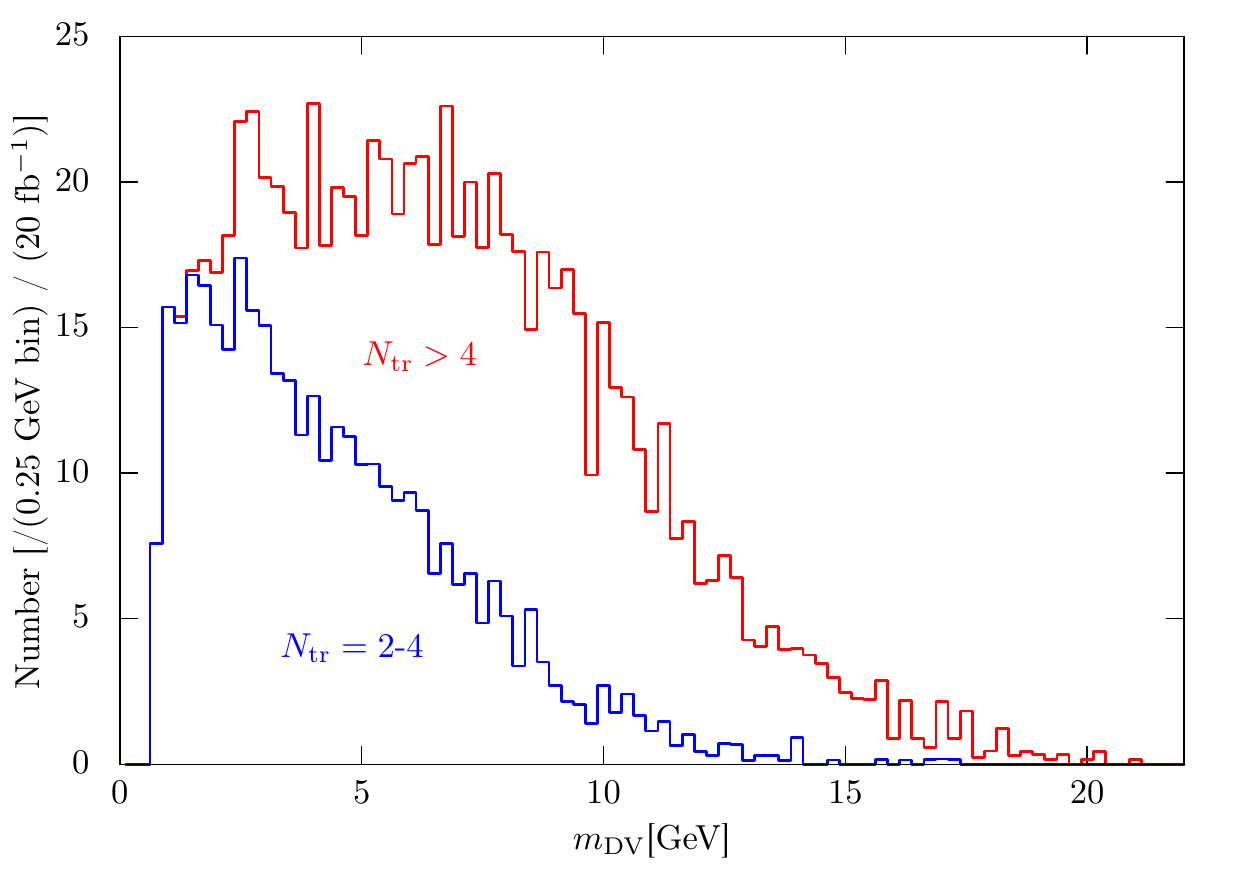}} 
\subcaptionbox{\label{eff} Single DV reconstruction
 efficiency}{\includegraphics[width=0.45\textwidth]{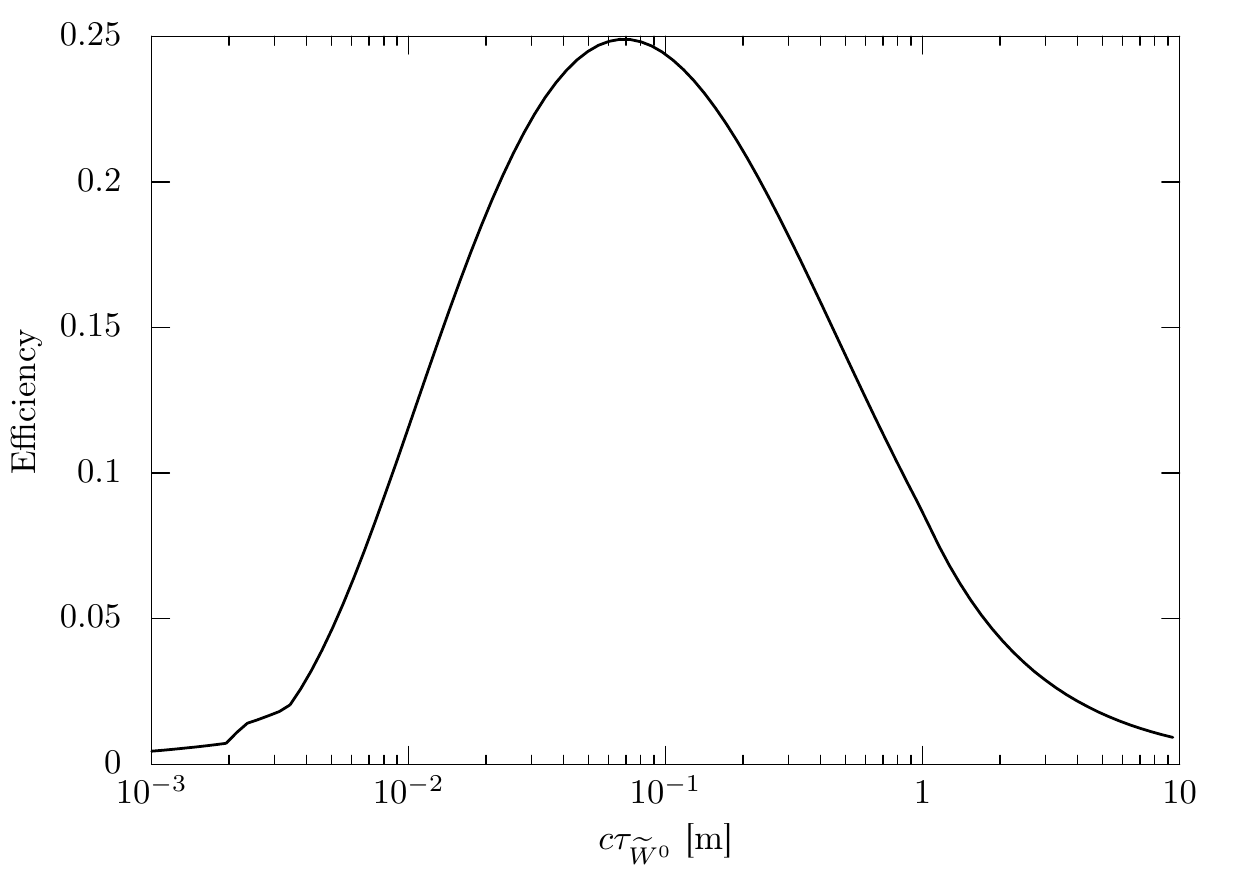}} 
\caption{
(a) Distributions of $m_{\rm DV}$. Here, we set
 $M_{\widetilde{W}}=400$~GeV and $M_{\widetilde{B}} = 370$~GeV. Winos
 are produced directly, which are assumed to decay through the Higgs
 exchange process.   (b) Single DV reconstruction efficiency as a
 function of the neutral wino decay length. 
}
\label{fig:cross}
\end{figure}

To see expected size of $m_{\rm DV}$, in Fig.~\ref{mdv}, we show the
distributions of $m_{\rm DV}$. Here we consider the wino direct
production case, with setting the wino and bino masses to be
$M_{\widetilde{W}}=400$~GeV and $M_{\widetilde{B}} = 370$~GeV,
respectively. A produced neutral wino is assumed to decay into a bino
via the Higgs exchange process. As can be seen, $m_{\rm DV}$ is expected
to be $\sim 10$~GeV, if $\Delta M = 30$~GeV, and a sizable number of
events provide $m_{\rm DV}>10$~GeV.

We also need the DV detection efficiency to estimate the number of
signals. We estimate it from the result of the ATLAS gluino search
\cite{Aad:2015rba}. Figure 19 of Ref.~\cite{Aad:2015rba} shows a
lower-bound on the gluino mass as a function of the gluino decay length
$c\tau_{\widetilde g}$, with the LSP mass of 100 GeV being assumed.
In this case, the DV+$E^{\rm{miss}}_{\rm{T}}$ search provides the
strongest constraints in most of the parameter region. The mass
difference between the gluino and the LSP is typically greater than
1~TeV in this plot. Since the transverse momenta of jets and the LSP
produced by the gluino decay are large, acceptance rates for the trigger
and the event filter are expected to be almost 100\%. Moreover, the
acceptance rate of the truth-level events for the requirements of the DV
criteria and the trackless jets is also expected to be almost 100\%. As
a consequence, we can estimate the decay-length dependent detection
efficiency of DVs, $\epsilon_{\widetilde{g}}$, as
\begin{align}
\epsilon_{\widetilde{g}} (c\tau) =  \frac{\sigma_{\rm obs} }{\sigma({ p p
 \to \widetilde g \widetilde g})}~, 
\end{align}
where $\sigma_{\rm obs} ~(= 0.15~{\rm fb})$ is observed upper-bound
on the cross section of the DV+$E^{\rm{miss}}_{\rm{T}}$ events
\cite{Aad:2015rba}. In this case, for each long-lived gluino DV event,
the number of DVs is two. Therefore, the reconstruction efficiency of a
single DV, $\epsilon_{\text{DV}}$, is about a half of
$\epsilon_{\widetilde{g}}$. For $c\tau_{\widetilde g} \gtrsim 1$~m, the
DV reconstruction efficiency would be inversely proportional to
decay length, since gluinos must decay within the tracker system. 
Although Fig.~19 of Ref.~\cite{Aad:2015rba} only shows the case of
$1<c\tau_{\widetilde g} <1000$~mm, we extrapolate the efficiency up to
$c\tau_{\widetilde g} > 1$~m taking into account a suppression factor
of $\propto (c\tau_{\widetilde g})^{-1}$. 
In Fig.~\ref{eff}, we show the estimated efficiency
$\epsilon_{\text{DV}}$ as a function of the decay length of the neutral
wino. This figure shows that the DV reconstruction efficiency is sizable
for $1~\text{cm}\lesssim c\tau_{\widetilde{W}^0} \lesssim 1$~m, which is
maximized when $c\tau_{\widetilde{W}^0} \sim 10$~cm. This estimation
assumes the velocity distribution of the long-lived winos to be the same
as that of long-lived gluinos in Ref.~\cite{Aad:2015rba}. Although these
two distributions are potentially different, we do not expect that this
difference alters our results drastically.

The total acceptance rate is the product of the efficiency  and the rate
of passing the above  $E^{\rm{miss}}_{\rm{T}}$ triggers and the DV
criteria. 
To estimate the acceptance rate, we use the program packages {\sc
Madgraph5} \cite{Alwall:2011uj}, {\sc Pythia6} \cite{Sjostrand:2006za},
and {\sc Delphes3} \cite{deFavereau:2013fsa}, while for the cross
sections of the SUSY particles we use {\sc Prospino2}
\cite{Beenakker:1996ed}. 

Note that the estimation of the prospects discussed below possibly
suffers from large uncertainties. At first, the DV detection efficiency
depends on the momentum distributions of the particles which generate
DVs. The small mass difference $\Delta M$ reduces the number of tracks
and makes their momenta lower, which may impair the DV reconstruction
efficiency. On the other hand, the current DV criteria is designed for
the high-mass DVs, and not optimized for the low-mass DVs like the
present wino case. A small modification in the DV criteria can
drastically change the acceptance rate as seen in Fig.~\ref{mdv} and can
potentially improve the detection efficiency. Moreover, although the
present DV reconstruction efficiency is obtained for DVs associated with
light-quark jets, we assume the neutral wino mainly decays into
$b$ jets in the following analysis. With the present ATLAS analysis, the
reconstruction efficiency of a DV accompanied with $b$ jets is expected
to be worse than those with light quarks. 
However, the neutral wino decaying into two $b$ quarks provides
characteristic signatures compared to light-quark decays; namely, the
$b$ quarks cause additional DVs originating from a DV given by a decay
of a long-lived neutral wino.
This specific signature may give further optimization of the search of
the wino DV and improve the reconstruction efficiency in the future LHC
experiments.
For more precise estimation, we need full detector simulation
and this is out of the scope of this paper. We use the above simplified
setup to see the prospects qualitatively in what follows.  

\subsection{LHC Prospects}
\label{sec:prospect}

First, let us discuss the case where winos are directly produced.
The crucial difference between the ATLAS gluino search and the present
direct wino search is the sizes of the missing transverse momentum and
the invariant mass of the tracks from DVs, due to the small mass
difference between the wino and bino. Both factors reduce the signal
acceptance of the wino processes. The missing energy mainly comes from
the back reaction of the initial state radiations in the wino
production. For $M_{\widetilde{W}} = 400$~GeV and $\Delta M = 30$~GeV, the
acceptance rates for the missing energy
($E^{\rm{miss}}_{\rm{T}}>100$~GeV for the 8~TeV running and 200~GeV for
14~TeV) and DV are about 3\% and 1\%, respectively. Here, we assume the
neutral wino decays into a pair of bottom quarks and a bino via the
Higgs boson exchange process.

\begin{figure}[t]
\centering
\subcaptionbox{\label{direct-wino}Direct wino production}{\includegraphics[width=0.45\textwidth]{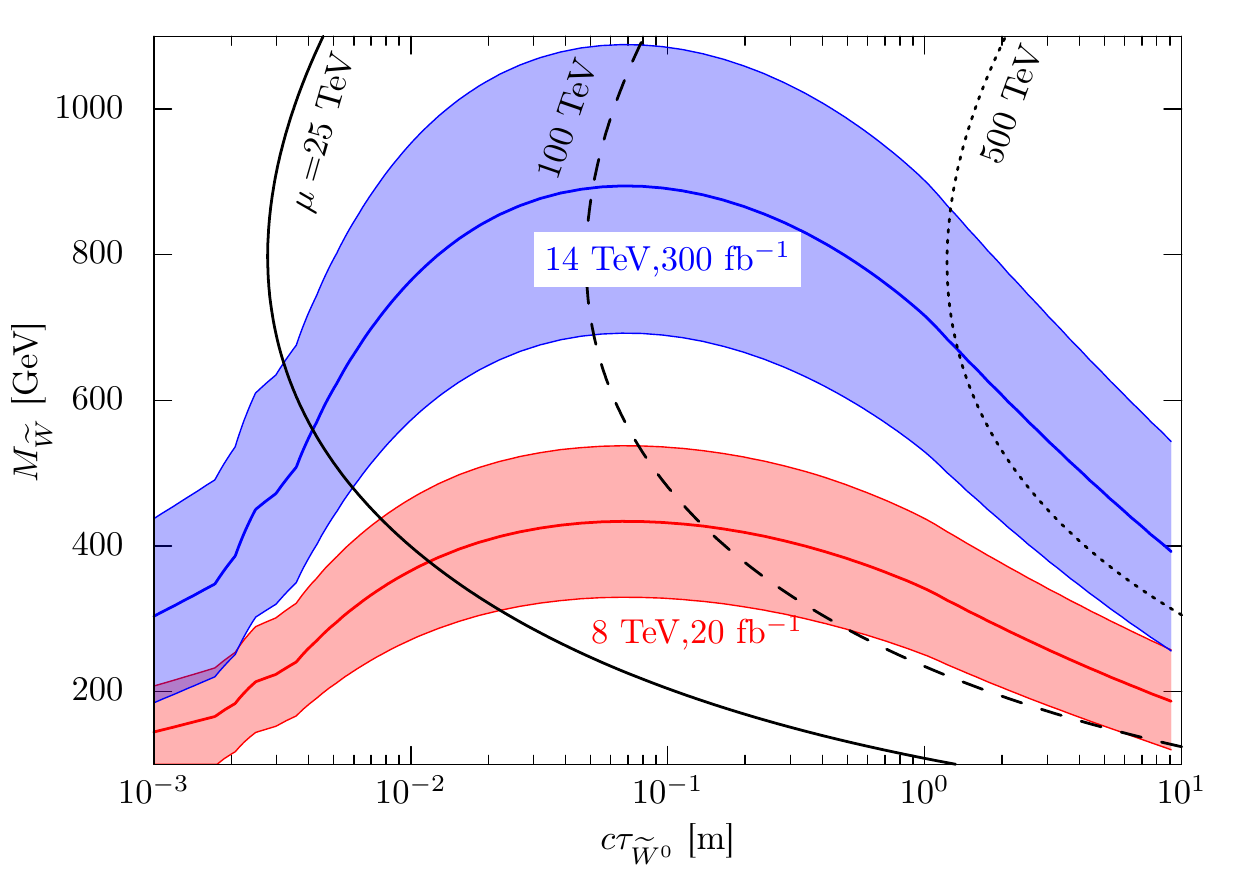}}
\subcaptionbox{\label{gluino}Gluino production}{\includegraphics[width=0.45\textwidth]{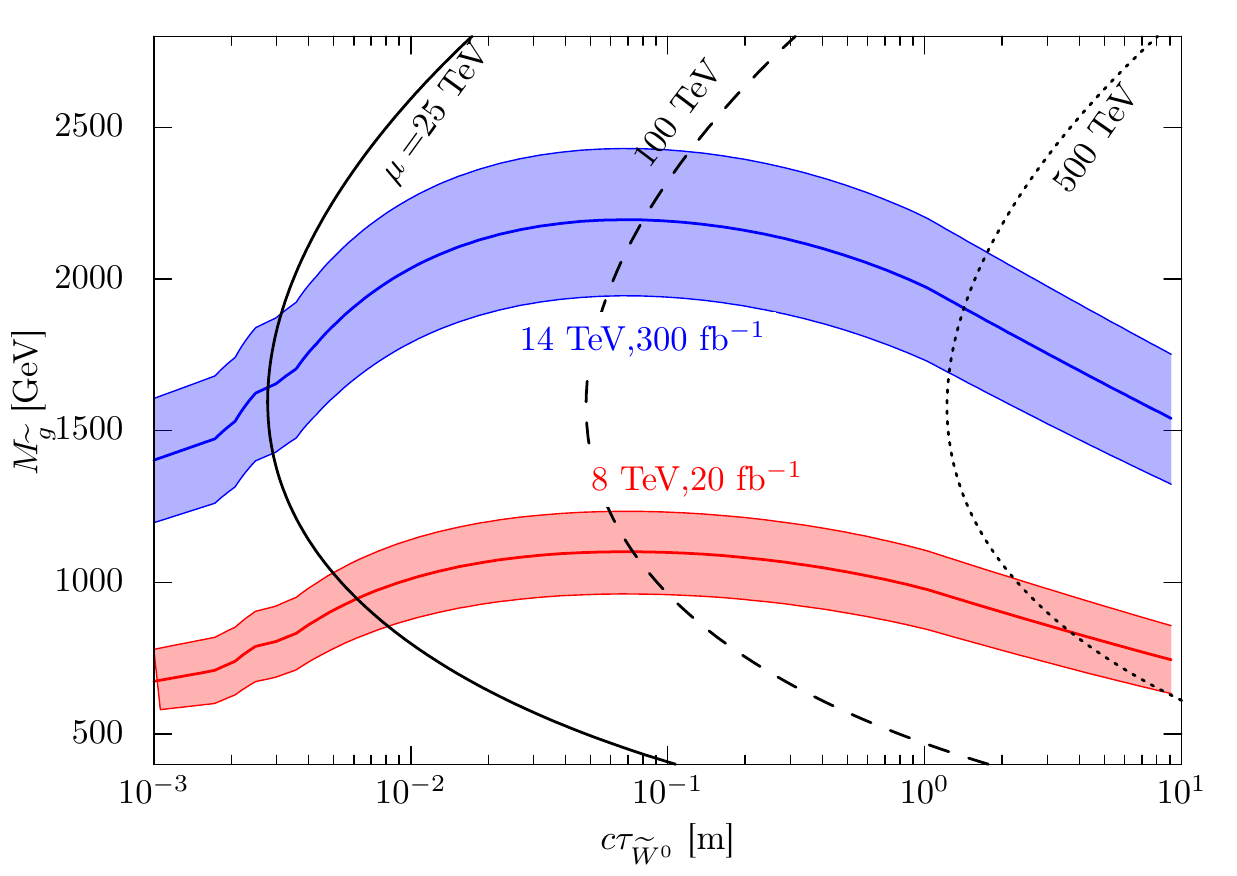}}

\caption{
Prospects for the long-lived wino search.
(a): direct wino production, (b): gluino production with $M_{\widetilde
 g}=2 M_{\widetilde W}$. In both cases, we assume the wino-bino mass
 difference is 30~GeV. Black solid, dashed, and dotted lines show
 contours corresponding to $\mu = 25$, 100, and 500~TeV, respectively,
 with $\tan\beta=2$ and $\Delta M$ taken so that it gives the correct DM
 abundance.  
}
\label{fig:prospect}
\end{figure}
In Fig.~\ref{direct-wino}, we show the prospects for the long-lived wino
search at the LHC. Here we assume zero background and require three
signal events. The mass difference $\Delta M$ is set to be 30~GeV.
The red and blue solid lines show the prospects for 8~TeV and 14~TeV LHC
run, respectively. Note that this estimation is based on our simplified
method described in the previous subsection. The DV reconstruct
efficiency in the real detector could be different, since the DV masses
in the current case are much smaller than those expected for the ATLAS
model points \cite{Aad:2015rba}. Moreover, $b$ jets from DVs may worsen
the efficiency. On the other hand, the current ATLAS analysis is not
optimized for the low-mass DV and $b$ jets, and thus future development
on search techniques for such DVs may improve the efficiency. These
possibilities result in large uncertainties in the present
estimation. Here, we estimate the uncertainties by scaling
the acceptance rate by factors of three and one third, and show them as
the bands in the figure.
For reference, we also show the decay length of the neutral wino for
$\mu=25$, 100, 500~TeV with $\tan\beta=2$ in the black solid, dashed,
and dotted lines, respectively, with $\Delta M$ taken so that it realizes
the correct DM abundance. From this figure, we find that using the DV
searches we may probe a wino with a mass of 400 GeV (800 GeV) at the 8
TeV (14 TeV) LHC if its decay length is ${\cal O}(10)$~cm. This should
be contrasted with other wino searches which do not use DVs. Without
DVs, the discovery of a wino having $\Delta M ={\cal O}(10)$~GeV is
extremely difficult; for such a neutral wino decaying into a bino via the
Higgs boson exchange, the soft leptons or mono-jet plus missing energy
searches at the LHC can provide essentially no constraints on this mass
spectrum.

Next, let us discuss the case in which winos are produced from decays of
gluinos. If gaugino masses are determined by only the anomaly mediation
effects and the threshold effects of the Higgsino loop, the gluino mass
is about 2--4 times larger than the wino mass in the case of the
bino-wino coannihilation. If the gluino mass is not so large, wino
productions through gluino decays are also active.
Unlike the direct wino production case, the acceptance rate of the
missing energy trigger is close to 100\% as long as the gluino-wino mass
difference is sufficiently large.
The rate for passing the above DV criteria is around 5\%.
The gluino decay mode strongly depends on the squark mass sector.
Here we assume that left-handed squarks dominantly contribute the gluino
decay, and the branching fraction of a gluino decaying into a wino,
$\widetilde{g} \to \widetilde{W}^0$, is about 30\%. In
Fig.~\ref{gluino}, we show the prospects for the wino search via this
channel. Here, we take $M_{\widetilde g}=2 M_{\widetilde W}$, and
$\Delta M = 30$~GeV. Again, the black solid, dashed, and dotted lines
show contours corresponding to $\mu = 25$, 100, and 500~TeV,
respectively, with $\tan\beta=2$ and $\Delta M$ taken so that it gives
the correct DM abundance. We find that in this case the 14~TeV LHC reach
for the gluino mass is as high as 2~TeV when $c\tau_{\widetilde{W}^0} \sim 10$~cm.

For the gluino-mediated case, a sizable mass difference between
gluino and wino leads to rather high jet activity and large missing
energy. Therefore, usual jets plus missing energy search (without DVs)
can probe a wide range of gluino masses.  For $M_{\widetilde{g}}=2
M_{\widetilde{W}}$, a gluino with a mass of 1 TeV (2 TeV) can be probed
with conventional jets plus missing energy search at the LHC8 (14). 
The gluino search with DVs discussed here may not drastically improve
the gluino discovery range, since the DV reconstruction efficiency is
quite low. Optimization of the DV criteria may increase this efficiency,
though. 
Even if gluinos are first discovered in the ordinary jets plus
missing energy searches, however, it is still important to search for
the gluino-mediated wino production events using DVs as well; it is
difficult for the former searches to distinguish the bino-wino
coannihilation scenario from merely the wino LSP case, while this is
possible for the DV searches. In addition, the DV search strategy may
allow us to extract the bino-wino mass difference through energy
measurements of decay product, which is quite important to test the
bino-wino coannihilation scenario.

\section{Conclusion and discussion}
\label{sec:conclusion}

In this paper, we study the neutral wino decay in the bino-wino
coannihilation scenario and discuss its collider signature. We find that
the neutral wino has a considerably long lifetime in the mini-split
spectrum, and is detectable at the LHC by means of the DV searches. 
To assess the prospects for the detectability, we study the direct
and gluino-mediated productions of the neutral wino at the 8 and 14~TeV
LHC running. It turns out that winos (gluinos) with a mass of 800~GeV
(2~TeV) can be probed at the next stage of the LHC in the former
(latter) production case. The search for the directly produced winos
with DVs is very powerful, compared to the conventional searches. For the gluino, this is not so
efficient and comparable to the conventional jets plus missing energy
searches. However, note that this conclusion can be altered since our
estimation of the detection rate has potentially large uncertainty. A
more realistic detector simulation is needed to obtain robust prospects.

In this paper, we focus on the parameter region where the tree-level
contributions are dominant. However, as discussed above, higher loop
processes may significantly contribute to the wino decay in certain
parameter region: a moderate $\mu$ case $|\mu|\lesssim {\cal O}(10)$~TeV,
a high-degeneracy mass case $\Delta M\sim 
{\cal O}(1)$~GeV, and so on. In these cases, the decay branches of wino
are altered, accordingly we may need to modify our collider search strategy. 
Detailed study with a more realistic detector simulation and more
accurate evaluation of the wino decay will be done elsewhere.

If the Higgsino mass is rather small, then the DV search strategy
discussed in this paper does not work as the neutral wino decay length
is too short, $c\tau_{\widetilde{W}^0}<1$~mm. However, it is possible to
probe such a region with low-energy precision experiments like the DM
direct detection experiments and the measurements of electric dipole
moments \cite{Nagata:2014wma, Hisano:2014kua, Nagata:2014aoa}. The
interplay between the DV search and these experiments will also be
discussed on another occasion.

Lastly let us comment on the prospects of future lepton colliders.
The high-energy lepton collider is very powerful tool to probe directly or indirectly the wino sector \cite{Gunion:2001fu,Harigaya:2015yaa}.
However it is non-trivial whether the lepton collider can observe the
wino DV signal, since the neutral wino production cross section is quite
small. Still, neutral winos may be produced through a loop process or
multi-particle production channel. In contrast, they can be
considerably produced both directly and indirectly at the LHC. In this sense,
the LHC experiment is quite suitable for the DV wino search.

\section*{Acknowledgments}

We thank Kazuki Sakurai for informing us of his related work in
Ref.~\cite{Rolbiecki:2015gsa}. The work of N.N. is supported by Research
Fellowships of the Japan Society for the Promotion of Science for Young
Scientists.

\bibliographystyle{aps}
\bibliography{ref}

\end{document}